\documentclass[preprint,12pt]{elsarticle}

\usepackage{graphicx}
\usepackage{amssymb}
\usepackage{color}

\journal{Optics Communications}
\begin{document}

\begin{frontmatter}

\title{Lensing and Waveguiding of Ultraslow Pulses in an Atomic Bose-Einstein Condensate}

\author[a]{Devrim Tarhan\corref{cor1}}

\ead{dtarhan@harran.edu.tr}

\author[b]{Alphan Sennaroglu}

\author[b,c]{\"{O}zg\"{u}r E. M\"{u}stecapl\i{}o\~{g}lu}

\address[a]{Department of Physics,
Harran University, Osmanbey Yerle\c{s}kesi, \c{S}anl\i{}urfa, 63300,
Turkey}

\address[b]{Department of Physics, Ko\c{c} University, Rumelifeneri yolu,
Sar\i{}yer, Istanbul, 34450, Turkey}

\address[c]{Institute of Quantum Electronics, ETH Zurich
Wolfgang-Pauli-Strasse 16, CH-8093 Zurich, Switzerland}

\cortext[cor1]{Corresponding author. Address:Department of
Physics, Harran University, Osmanbey Yerle\c{s}kesi,
\c{S}anl\i{}urfa, 63300, Turkey. Tel.: +90 414 3183578; fax: +90
414 3440051.}

\begin{abstract}

We investigate lensing and waveguiding properties of an atomic
Bose-Einstein condensate for ultraslow pulse generated by
electromagnetically induced transparency method. We show that a
significant time delay can be controllably introduced between the
lensed and guided components of the ultraslow pulse. In addition,
we present how the number of guided modes supported by the
condensate and the focal length can be controlled by the trap
parameters or temperature.

\end{abstract}


\end{frontmatter}

\newpage
\section{Introduction}
\label{sec:intro}

Quantum interference effects, such as electromagnetically induced
transparency (EIT) \cite{harris,atac}, can produce considerable
changes in the optical properties of matter and have been utilized
to demonstrate ultraslow light propagation through an atomic
Bose-Einstein condensate (BEC) \cite{hau}. This has promised a
variety of new and appealing applications in coherent optical
information storage as well as in quantum information processing .
However, the potential of information storage in such systems is
shadowed by their inherently low data rates. To overcome this
challenge, exploitation of transverse directions for a multimode
optical memory via three dimensional waveguiding of slow EIT pulse
\cite{cheng} has been recently suggested for BECs \cite{tarhan}.
Transverse confinement of slow light is also quintessential for
various proposals of high performance intracavity and nanofiber
slow light schemes (See e.g. Ref. \cite{fam} and references
therein). Furthermore, temperature dependence of group velocity of
ultraslow light in a cold gas has been investigated for an
interacting Bose gases \cite{morigi}.

A recent experiment, on the other hand, has drawn attention that
ultracold atomic systems with graded index profiles may not
necessarily have perfect transverse confinement due to
simultaneously competing effects of lensing and waveguiding
\cite{venga1}. The experiment is based upon a recoil-induced
resonance (RIR) in the high gain regime, employed for an ultracold
atomic system as a graded index waveguiding medium. As a result of
large core radius with high refractive index contrast, and strong
dispersion due to RIR, radially confined multimode slow light
propagation has been realized \cite{venga1}. As also noted in the
experiment, a promising and intriguing regime would have few modes
where guided nonlinear optical phenomena could happen
\cite{venga1}.

It has already been shown that the few mode regime of ultraslow
waveguiding can be accessed by taking advantage of the sharp
density profile of BEC and the strong dispersion provided by the
usual EIT \cite{tarhan}. Present work aims to reconsider this
result by taking into account the simultaneous lensing component.
On one hand, the lensing could be imagined as a disadvantage
against reliable high capacity quantum memory applications. Our
investigations do aid to comprehend the conditions of efficient
transverse confinement. On the other hand, we argue that because
the lensing component is also strongly delayed with a time scale
that can be observably large relative to the waveguiding modes,
such spatially resolved slow pulse splitting can offer intriguing
possibilities for creating and manipulating flying bits of
information, especially in the nonlinear regime. Indeed, earlier
proposals to split ultraslow pulses in some degrees of freedom
(typically polarization), face many challenges of complicated
multi-level schemes, multi-EIT windows, and high external fields
\cite{agarwal}. Quite recently birefringent lensing in atomic
gaseous media in EIT setting has been discussed \cite{zhang}. The
proposed splitting of lensing and guiding modes is both
intuitively and technically clear, and easy to implement in EIT
setting, analogous to the RIR experiment.

The paper is organized as follows: After describing our model
system briefly in Sec. \ref{sec:model}, EIT scheme for an
interacting BEC is presented in Sec. \ref{sec:eit}. Subsequently
we focus on the lensing effect of the ultraslow pulse while
reviewing already known waveguiding results shortly in Sec.
\ref{sec:propagation}. Main results and their discussion are in
Sec. \ref{sec:results}. Finally, we conclude in Section
\ref{sec:concl}.


\section{Model System}
\label{sec:model}

\begin{figure}[htbp]
\centerline{\includegraphics[width=8.3cm]{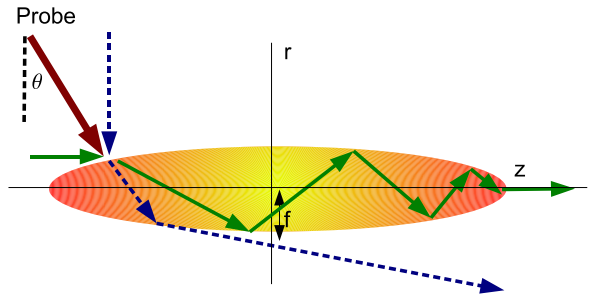}}
\caption{Lensing and waveguiding effects in a slow Gaussian beam
scheme with an ultracold atomic system.} \label{fig1}
\end{figure}

We consider oblique incidence of a Gaussian beam pulse onto the
cigar shaped condensate as depicted in Fig. (\ref{fig1}). Due to
particular shape of the condensate two fractions, the one along
the long ($z$) axis of the condensate and the parallel the short
($r$) axis, of the incident Gaussian beam exhibit different
propagation characteistics. The $r$-fraction exhibit the lensing
effect and focused at a focal length ($f$) while the axial
component would be guided in a multi mode or single mode
formation.

The angle of incidence ($\theta$) controls the fraction of the
probe power converted either to the lensing mode or to the guiding
modes. When both lensing and waveguiding simultaneously happen in
an ultraslow pulse propagation set up, an intriguing possibility
arises. Different density profiles along radial and axial
directions translates to different time delays of the focused and
guided modes. Due to significant difference in the optical path
lengths of guided and focused components, an adjustable relative
time delay can be generated between these two components. As a
result, these two components become spatially separated. The
fraction of the beam parallel to the short axis of the condensate
undergoes propagation in a lens-like quadratic index medium
resulting in a change in the spot size or the beam waist of the
output beam. We call this as the lensing effect and the
corresponding fraction as the lensed-fraction. The fraction of the
beam propagating along the long axis is propagating in
weakly-guided regime of graded index medium that can be described
in terms of LP modes. This is called as guiding effect and the
corresponding fraction is called as guided fraction. We use the
usual Gaussian beam transformation methods under paraxial
approximation to estimate the focal length. In addition our aim is
to estimate time delay between these two fractions. The output
lensed and guided fractions of the incident beam are delayed in
time relative to each other. In general temporal splitting depends
on modal, material and waveguide dispersions. For a simple
estimation of relative time delay of these components, we consider
only the lowest order modes in the lensed and the guided
fractions, and take the optical paths as the effective lengths of
the corresponding short and long axes of the condensate. In this
case we ignore the small contributions of modal and waveguide
dispersions and determine the group velocity, same for both
fractions, by assuming a constant peak density of the condensate
in the material dispersion relation.


\section{EIT scheme for an interacting BEC}
\label{sec:eit}

A Bose gas can be taken as condensate part and thermal part at low
temperature. Following Ref. \cite{naraschewski}, density profile
of BEC can be written by $\rho(\vec{r})=\rho_c(\vec{r})
+\rho_{{\rm th}}(\vec{r})$, where $\rho_{{\rm c}}(\vec{r})=
[(\mu(T)-V(\vec{r}))/U_0] \Theta(\mu-V(\vec{r}))$ is the density
of the condensed atoms and $\rho_{{\rm th}}$ is the density of the
thermal ideal Bose gas. Here $U_0=4\pi\hbar^2 a_{s}/m$; $m$ is
atomic mass; $a_s$ is the atomic s-wave scattering length.
$\Theta(.)$ is the Heaviside step function and $T_C$ is the
critical temperature. The external trapping potential is
$V(\vec{r})=(m/2) (\omega_r^2 r^2+\omega_z^2 z^2)$ with trap
frequencies $\omega_{r},\omega_{z}$  for the radial and axial
directions, respectively. At temperatures below $T_c$, the
chemical potential $\mu$ is evaluated by
$\mu(T)=\mu_{TF}(N_0/N)^{2/5}$, where $\mu_{TF}$ is the chemical
potential obtained under Thomas-Fermi approximation,
$\mu_{TF}=((\hbar\omega_t)/2)(15Na_s/a_h)^{2/5}$, with $\omega_t =
(\omega_z\omega_r^2)^{1/3}$  and
$a_h=\sqrt{\hbar/(\omega_z\omega_r^2)^{1/3}}$, the average
harmonic oscillator length scale. The condensate fraction is given
by $N_0/N=1-x^3-s \zeta(2) / \zeta(3)x^2(1-x^3)^{2/5}$, with
$x=T/T_c$, and $\zeta$ is the Riemann-Zeta function. The scaling
parameter $s$ is given by $s=\mu_{TF}/k_BT_C=(1/2)\zeta(3)^{1/3}
(15N^{1/6}a_s/a_h)^{2/5}$.

Treating condensate in equilibrium and under Thomas-Fermi
approximation (TFA) is common in ultraslow light literature and
generally a good approximation because density of an ultracold
atomic medium is slowly changing during the weak probe
propagation. The propagation is in the order of microseconds while
atomic dynamics is in millisecond time scales. Due to weak probe
propagation under EIT conditions, most of the atoms remain in the
lowest state\cite{dutton}. The validity of TFA further depends on
the length scale of the harmonic potential. If the length scale is
much larger than the healing length, TFA with the harmonic
potential works fine. The healing length of the BEC is defined as
$\xi =[1/(8\pi n a_s)]^{1/2}$ \cite{pethick} where $n$ is the
density of an atomic Bose-Einstein condensate and it can be taken
as $n=\rho(0,0)$. We consider range of parameters in this work
within the range of validity of TFA. The interaction of atomic BEC
with strong probe and the coupling pump field may drive atomic BEC
out of equilibrium. This non- equilibrium discussion is beyond the
scope of the present paper.

We consider, beside the probe pulse, there is a relatively strong
coupling field interacting with the condensate atoms in a
$\Lambda$-type three level scheme with Rabi frequency $\Omega_c$.
The upper level is coupled to the each level of the lower doublet
either by probe or coupling field transitions. Under the weak
probe condition, susceptibility $\chi$ for the probe transition
can be calculated as a linear response as most of the atoms remain
in the lowest state. Assuming local density approximation,
neglecting local field, multiple scattering and quantum
corrections and employing steady state analysis we find the
well-known EIT susceptibility \cite{harris,eit3},
$\chi_{i},i=r,z$, for either radial ($r$) or axial ($z$) fraction
of the probe pulse. Total EIT susceptibility for BEC in  terms of
the density $\rho$ can be expressed as $\chi_{i}=\rho_i \,\chi_1$
in the framework of local density approximation. Here $\chi_1$ is
the single atom response given by
\begin{eqnarray}
\label{eit} \chi_1  = \frac{{ \left| {\mu } \right|^2
}}{{\varepsilon _0 \hbar }} \frac{{i(-i\Delta  + \Gamma _2
/2)}}{{{(\Gamma _2 /2 - i\Delta ) (\Gamma _3 /2 - i\Delta ) +
\Omega _C ^2 /4} }},
\end{eqnarray}
where $\Delta$ is the detuning from the resonant probe transition
. For the ultracold atoms and assuming co-propagating laser beams,
Doppler shift in the detuning is neglected. $\mu$ is the dipole
matrix element for the probe transition. It can also be expressed
in terms of resonant wavelength $\lambda$ of the probe transition
via $ \mu  = 3\varepsilon _0 \hbar \lambda^2 \gamma /8\pi ^2 $,
where $\gamma$ is the radiation decay rate of the upper level.
$\Gamma_2$ and $\Gamma_3$ denote the dephasing rates of the atomic
coherences of the lower doublet. At the probe resonance, imaginary
part of $\chi$ becomes negligible and results in turning an
optically opaque medium transparent.


\section{Propagation of beam through a quadratic index medium}
\label{sec:propagation}
\subsection{Lensing effect}
\label{sec:lens}

We can neglect thermal part of a Bose gas due to the high index contrast between
 the condensate and the thermal gas background so that $\rho=\rho_c$.
 We specifically consider a gas of $N=8\times10^6$
$^{23}$Na atoms with $\Gamma_3=0.5\gamma$, $\gamma=2\pi\times
10^7$Hz, $\Gamma_2=7\times 10^3$ Hz, and $\Omega_c=2\gamma$. We
take $\omega_{r}=160$ Hz and $\omega_{z}=40$ Hz. For these
parameters, we evaluate $\chi^{\prime}=0.02$ and
$\chi^{\prime\prime}=0.0004$ at $\Delta=0.1\gamma$, where
$\chi^{\prime}$ and $\chi^{\prime\prime}$ are the real and
imaginary parts of $\chi$, respectively. Neglecting
$\chi^{\prime\prime}$, the refractive index becomes
$n=\sqrt{1+\chi^{\prime}}$. In the $z$ direction it can be written
as \cite{Yariv,Alphan}
\begin{eqnarray}
\label{refindex} n(z)=n_0[1-\beta_z^2 z^2]^{1/2},
\end{eqnarray}
where $n_0=(1+\mu\chi_1^{\prime}/U_0)^{1/2}$ and quadratic index
coefficient is $\beta_z^2=\chi_1^{\prime} m \omega_z^2/(2U_0
n_0^2)$. Thomas-Fermi radius for the axial coordinate is given by
$R_{TF_{z}}=\sqrt{2\mu(T)/m\omega_z^2}$. Expanding Eq.
(\ref{refindex}) in Taylor series, the refractive index reduces to
\begin{eqnarray}
\label{refindex1} n(z)\approx n_0[1-\frac{1}{2}\beta_z^2 z^2].
\end{eqnarray}
\begin{figure}[htbp]
\centering{\vspace{0.5cm}}
\includegraphics[width=8cm]{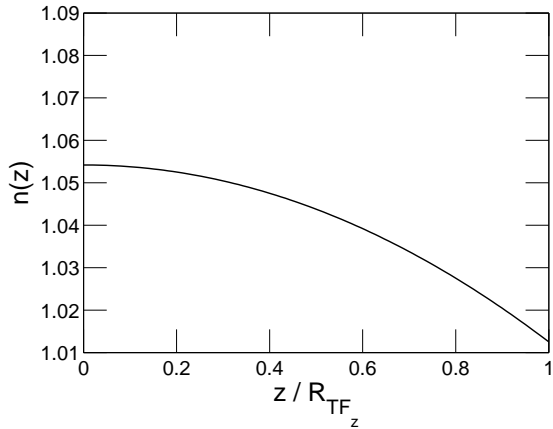}
\caption{$z$ dependence of the refractive index of a condensate of
$8 \times 10^6$ $^{23}$ Na atoms at $T=296$ nK under off-resonant
EIT scheme. The other parameters used are $\Omega_c=1.5\times
\gamma$, $\Delta=0.1\times \gamma$, $M=23$ amu, $\lambda_0 = 589$
nm, $\gamma = 2\pi\times 10^7$ Hz, $\Gamma_3=0.5\gamma$,
$\Gamma_2=7 \times 10^3$ Hz.} \label{fig2}
\end{figure}

The refractive index as a function of $z$ is shown in Fig.
\ref{fig2}. With such a refractive index profile, the atomic
medium can act as a thin lens for the component of the probe in
the radial direction. In the case of a medium with a quadratic
index variation, we can estimate the focal length by using
geometrical or beam optics \cite{Yariv,Alphan}. In the paraxial
geometrical optics regime, where the angle made by the beam ray
with optic axis is small, the differential equation satisfied by
the ray height s(r) is $d^2 s / dr^2 + \beta_z^2 r=0$
\cite{Yariv,Alphan}. The initial ray height and initial slope are
$r_i$ , $\theta_i = {ds / dr}|_i$ respectively. The solutions of
the differential equation are $s(r)=r_i \cos(\beta_z r)+1 /
\beta_z \sin(\beta_z r) \theta_i$ and $\theta(z)= ds/dr= -\beta_z
\sin(\beta_z r)r_i+\cos(\beta_z r)\theta_i$. Hence, the ray
transformation matrix $M_T$ which connects the initial and final
ray vectors is given by Eq. (\ref{mt}), where $r$ is the
propagation length in the medium. \cite{Yariv,Alphan,Yariv1}
\begin{eqnarray}
\label{mt} M_T = \left( \begin{array}{cc} \cos(\beta_z r) &
\frac{1}{\beta_z} \sin(\beta_z r)
\\-\beta_z \sin(\beta_z r)& \cos(\beta_z r) \end{array} \right).
\end{eqnarray}

The same ray transformation matrix given by Eq. (\ref{mt}) can be
used to investigate the effect of the quadratic medium on the
light in the beam optics regime. Consider a collimated
Gaussian beam incident on a focusing ultra cold medium represented
by $M_T$. The $q$ parameter $q(r)$ of the Gaussian beam which
describes the spot size and the radius of curvature of the beam is
given by
\begin{eqnarray}
\label{radcurv} \frac{1}{q(r)}=\frac{1}{R(r)} - i \frac{\lambda}{n
\pi w(r)^2},
\end{eqnarray}
where $w(r)$ is the spot-size function, and n is the medium
refractive index.  If the incident beam is collimated,
$R(r)\rightarrow\infty$ , and the initial $q$-parameter $q_1$
becomes $i z_0$, where the Rayleigh range $z_0$ is given by $z_0=
\pi\omega_0^2/\lambda$ .  Here, $w_0$ is the initial beam waist on
the BEC and we assumed that the background refractive index is
nearly $1$. After a distance of $r$ inside the BEC, the
transformed $q$-parameter $q_2$ will given by
\begin{eqnarray}
\label{output} q_2=\frac{iz_0\cos(\beta_r r)+\frac{1}{\beta_z}
\sin(\beta_z r)}{-iz_0\beta_z \sin(\beta_z r)  + \cos(\beta_z r)}.
\end{eqnarray}

We have two cases to consider for focusing.  If the BEC is very
thin, the focal length will be approximately given by $1/\beta_z^2
L_r$ where $L_r$ is the transverse width of the BEC.  If the BEC
length is not negligible, the next collimated beam will be formed
inside the BEC at the location $L_f$, given by $\beta_z L_f = \pi
/2$.
\subsection{Waveguiding effect}
\label{sec:guide}
Along the radial direction, similar to previous
treatment, refractive index profile in the radial direction can be
written as
\begin{eqnarray}
\label{refindexr} n(r) = \cases{n_0 [1-\frac{1}{2} \, \beta_r^2 \,
r^2]^{1/2} & $r\leq R_{TF_{r}}$. \cr 1 & $r\geq R_{TF_{r}}$},
\end{eqnarray}
where $n_0=(1+\mu\chi_i^{\prime}/U_0)^{1/2}$ and
$\beta_r^2=\chi_1^{\prime} m \omega_z^2/(2U_0 n_0^2)$.
Thomas-Fermi radius is given by
$R_{TF_{r}}=\sqrt{2\mu(T)/m\omega_r^2}$.
\begin{figure}[htbp]
\centering{\vspace{0.5cm}}
\includegraphics[width=8cm]{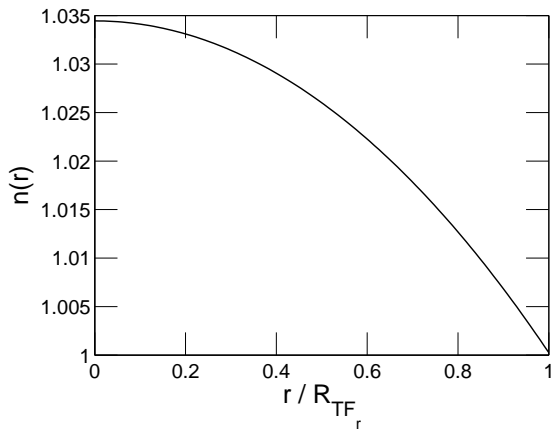}
\caption{$r$ dependence of the refractive index of the condensate.
The parameters are the same as Fig. (\ref{fig2})} \label{fig3}
\end{figure}

The refractive index which is given in Eq.(\ref{refindexr}) is
plotted as a function of $r$ in Fig.\ref{fig3}. Expanding Eq.
(\ref{refindexr}) in Taylor series, the refractive index reduces
to $n(r)\approx n_0(1-0.5\beta_r^2 r^2)$. Such an index behavior
is analogous to the one of a graded index fiber \cite{tarhan}.
Slow light propagation through the condensate can be described
similar to that of the weakly guided regime of the graded index
fiber, where the optical modes can be given in terms of linearly
polarized modes (LP modes). The mode profiles are determined by
solving the wave equation which reduces to the Helmholtz equation
\cite{tarhan}.

We use the cylindrical coordinates as the refractive index $n(r)$
is axially symmetric. The wave equation for the axial fraction of
the probe field is $[\nabla^2+k^2]E=0$, where $k^2=k_z^2 + k_r^2$
and $\nabla^2$ is the Laplacian operator in cylindrical
coordinate. Here $k_r$ is the radial wave number and $k_z$ is the
propagation constant in the $z$ direction. The solution for the
wave equation is $E=\psi(r) \cos(l\phi)\exp[i(\omega t-k_z z)]$.
Here $l=0,1,2,3,...$ and $\phi$ is the absolute phase. If we put
this solution in the wave equation, we get the Helmholtz radial
equation
\begin{eqnarray}
 [d^2/dr^2+(1/r) d/dr+p^2(r)]\psi(r)=0,
 \label{Helmholtz}
\end{eqnarray}
in which $p^2(r)=(k_{0}^2 n^2(r)-k_z ^2-l^2/r^2)$ \cite{Yariv}.
Here $k_0$ can be expressed in terms of resonant wavelength
$\lambda$ of the probe transition
$k_0=2\pi/\lambda=1.07\times10^7$ (1/m).

We use transfer matrix method developed in Ref. \cite{tarhan} in
order to solve the Helmholtz equation. Eq.(\ref{Helmholtz}).

We assume that the atomic cloud can be described by layers of
constant refractive index, such that, their indices monotonically
increase towards the center of the cloud. By taking sufficiently
large number of thin layers such a discrete model can represent
the true behavior of the refractive index. For constant index,
analytical solutions of the Helmholtz equation can be found in
terms of Bessel functions. Such solutions are then matched at the
shell boundaries of the layers. Doing this for all the layers,
electromagnetic boundary conditions provide a recurrence relation
for the Bessel function coefficients, whose solution yields the
wave number $k$. The mode profiles determined by this method are
shown in Fig. (\ref{fig6}), and Fig. (\ref{fig7}).


\section{Results and Discussions}
\label{sec:results}
\subsection{Lensing Properties}
\label{sec:lensresult}

The condensate can act as a lens or a guiding medium depending on
the Rayleigh range of the probe relative to the effective length
of the condensate \cite{venga1}. For the cigar shaped BEC geometry
we consider, the radial fraction of the probe is subject to
lensing while the axial fraction is guided. Let us first examine
the focal length of the condensate for the radial fraction of the
incident probe. For that aim we need to determine the effective
radial length
\begin{eqnarray}
\label{eq:radial_size} L_r=\left[\int_V d^3 r
r^2\rho(r,z)\right]^{1/2},
\end{eqnarray}
corresponding to the radial rms width of the density distribution.
Radial component of the probe field enters the medium and
converges at a focus distance, determined
\begin{equation}
f=\frac{1}{\beta_z^{2} L_r}.
\end{equation}
\begin{figure}[htbp]
\centering{\vspace{0.5cm}}
\includegraphics[width=8cm]{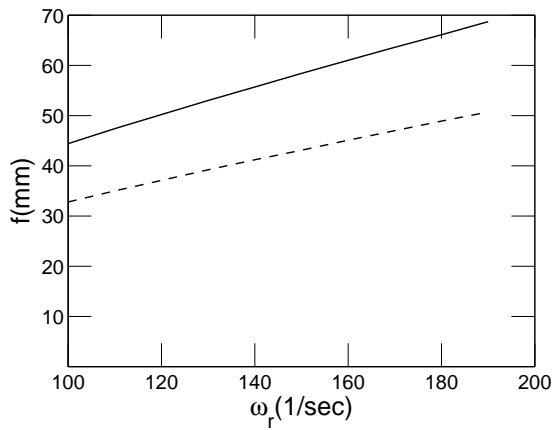}
\caption{Behavior of the focus length with respect to trap
frequency for the radial direction and scattering length. The
solid curve is for $a_{s}=5$ nm and the dashed curve is for
$a_{s}=7$ nm. The other parameters are the same as Fig.
(\ref{fig2}) } \label{fig4}
\end{figure}

Focal length is plotted in  as shown in Fig. (\ref{fig4}) as a function of
trap frequency for the radial direction and with the scattering length at a
temperature $T=296$ nK. The increase in the radial trap
frequency leads to decrease in the radial size of the BEC,
which causes the rise of the focal length seen in Fig. (\ref{fig4}). In contrast,
the size of the condensate increases by atom-atom interactions,
characterized by the scattering length
$a_s$, which causes an overall reduction of the focal length.

\subsection{Waveguiding Properties}
\label{sec:guideresults}

For the cigar shaped BEC geometry we consider again the axial
fraction of the probe is subject to guided. The effective axial
length is determined by
\begin{eqnarray}
\label{eq:axial_size}
L_z=\left[\frac{4\pi}{N}\int_0^\infty\,r\mathrm{d}r\int_0^\infty\,\mathrm{d}z
z^2\rho(r,z)\right]^{1/2}.
\end{eqnarray}
The axial length $L_z$ is an effective length corresponding to the
axial width of the density distribution. Group velocities
($v_{gi},i=r,z$) of the different density profiles for both radial
and axial directions can be calculated from the susceptibility
using the relations
\begin{eqnarray}
\frac{1}{{v_{gi} }} &=& \frac{1}{c} + \frac{\pi }{\lambda
}\frac{{\partial \chi _i }} {{\partial \Delta }}.\label{eq:vg}
\end{eqnarray}
Here, c is the speed of light and imaginary part of the
susceptibility is negligibly small relative to the real part of
it. EIT can be used to achieve ultraslow light velocities, owing
to the steep dispersion of the EIT susceptibility $\chi$
\cite{hau}. In general temporal splitting depends on modal,
material and waveguide dispersions. For a simple estimation of
relative time delay of these components, we consider only the
lowest order modes in the lensed and the guided fractions, and
take the optical paths as the effective lengths of the
corresponding short and long axes of the condensate. In this case
we ignore the small contributions of modal and waveguide
dispersions and determine the group velocity, same for both
fractions, by assuming a constant peak density of the condensate
in the material dispersion relation. Density profiles along radial
and axial directions lead to different time delays of the focused
and guided modes. Due to significant difference in the optical
path lengths of guided and focused components, a certain time
delay can be generated between these two components.
\begin{figure}[htbp]
\centering{\vspace{0.5cm}}
\includegraphics[width=8cm]{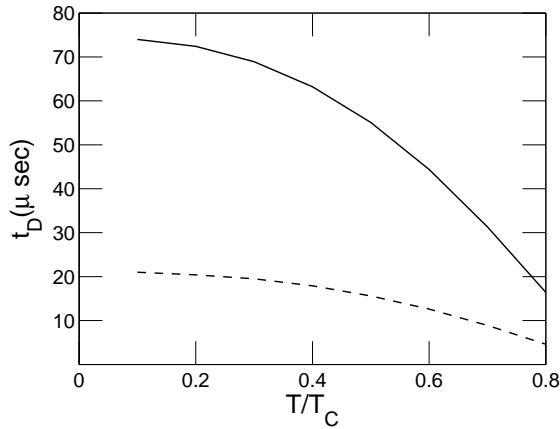}
\caption{Thermal behavior of the time delays of the guided and
focused fractions. The solid and dashed lines are for $t_{Dz}$ and
$t_{Dr}$ , respectively. The parameters are the same as Fig.
(\ref{fig2}) except $\Omega_c=0.5\gamma$} \label{fig5}
\end{figure}

Taking into account transverse confinement, ultraslow Gaussian
beam propagation along the long axis of a BEC can be described in
terms of multiple LP modes, propagating at different ultraslow
speeds \cite{tarhan}. In the geometry and system we consider here,
for instance, there can be found approximately $44$ modes at
temperature $T=42$ nK. We compare the delay time of the slowest
mode (the lowest guided mode) with that of the lensing mode which
can be respectively calculated by $t_{Dz}=L_z/v_{gz}$ and
$t_{Dr}=L_r/v_{gr}$. Thermal behaviors of the time delays are
shown in Fig. (\ref{fig5}). Fig. (\ref{fig5}) indicates that due
to different refractive indices seen by the axial and radial
fractions of the probe, significant temporal delay, $\sim
50\,\mu$s, can occur between them at low temperatures $T\sim 42$
nK. Time delays for each fraction decreases with the temperature.
Besides, the relative difference of their time delays diminishes
with the temperature. The condensed cloud shrinks due to the
increase in temperature, so that effective lengths of the BEC
diminish. Therefore, just below $T_C$, time delays drop to zero.

\begin{figure}[htbp]
\centering{\vspace{0.5cm}}
\includegraphics[width=3.5in]{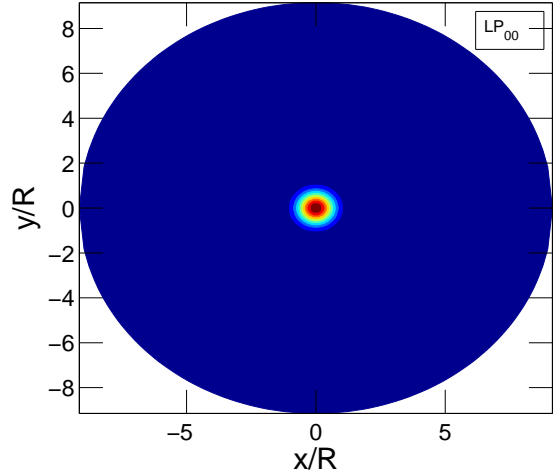}
\caption{Contour plot of intensity of the $\psi_{00} \exp[i(\omega
t-\beta_z z)]$ (LP$_{00}$ mode). The parameters are the same as
Fig. (\ref{fig5})} \label{fig6}
\end{figure}
\begin{figure}[htbp]
\centering{\vspace{0.5cm}}
\includegraphics[width=3.5in]{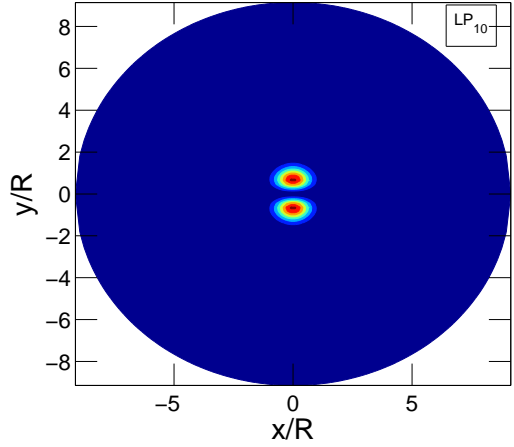}
\caption{Contour plot of intensity of the $\psi_{10} \cos(
\phi)\exp[i(\omega t-\beta_z z)]$ (LP$_{10}$ mode). The parameters
are the same as Fig. \ref{fig5}.} \label{fig7}
\end{figure}

At low temperatures waveguiding modes can occur in an atomic
Bose-Einstein condensate because  of an atomic Bose-Einstein
condensate. Translation of the Thomas-Fermi density profile of the
condensate to the refractive index makes the medium gain
waveguiding characteristics analogous to those of a graded index
fiber \cite{tarhan,venga1}. For the same set of parameters, for
which the radial fraction of the probe undergoes lensing effect,
the axial fraction is guided in multiple LP modes, as the
corresponding Rayleigh range is larger than the effective axial
length. The lowest two LP modes are shown in Fig. (\ref{fig6}),
and Fig. (\ref{fig7}). The total number of modes that can be
supported by the condensate is determined by the dimensionless
normalized frequency $V$ which is defined as
$V=(\omega/c)R(n_0^2-1)^{1/2}$ where
$R=\sqrt{2\mu(T)/m\omega_r^2}$ \cite{tarhan}. The radius of a
condensate that would support only single mode, LP$_{00}$, is
found to be $R \sim 1\,\mu$m. In principle this suggest that
simultaneous lensing and single LP mode guiding should also be
possible.
\newpage
%


\section{Conclusion}
\label{sec:concl}

We investigate simultaneous lensing and ultraslow waveguiding
properties of atomic condensate under EIT conditions by using a
quadratic-index model of the medium. The focus length, relative
time delay, and multiple guided mode characteristics are
determined taking into account three dimensional nature of the
system. In particular, dependence of focus length on atom-atom
interactions via s-wave scattering length, and on trap frequency
and temperature are examined. Our results reveal how to select a
suitable set of system parameters to tune Rayleigh range and the
aspect ratio of the cloud to make either lensing or guiding more
favorable along a particular direction. We have shown that the
focus length can be calibrated by the transverse trap frequency,
incoming wave length, scattering length and temperature. In
addition, time-delayed splitting of ultraslow Gaussian beam into
radial and axial fractions is found.

\section*{Acknowledgements}
We thank Z. Dutton for valuable and useful discussions. D.T.
was supported by TUBITAK-Kariyer grant No. 109T686. \"{O}.E.M.
acknowledges support by  TUBITAK (109T267) and DPT-UEKAE quantum cryptology
center.



\begin{thebibliography}{00}

%
\bibitem{harris}
S.E. Harris, Physics Today 50 (1997) 36-42.
%
\bibitem{atac}
M. Fleischhauer, A. Imamoglu, and J. P. Marangos,
Rev. Mod. Phys. 77 (2005) 633.
%
\bibitem{hau}
L.V. Hau, S.E. Harris, Z. Dutton, C.H. Behroozi, Nature 397 (1999) 594-598.
%
\bibitem{cheng}
J. Cheng, S. Han, Y. Yan, Phys. Rev. A 72 (2005) 021801(R).
%
\bibitem{tarhan}
D. Tarhan, N. Postacioglu, \"{O}. E. M\"{u}stecapl\i{}o\~{g}lu,
Opt. Lett. 32 (2007) 1038.
%
\bibitem{fam}
F. L. Kien, and K. Hakuta, Phys. Rev. A 79 (2009) 043813.
%
\bibitem{morigi}
G. Morigi, and G. S. Agarwal, Phys. Rev. A 79 (2000) 013801.
%
\bibitem{venga1}
M. Vengalattore, and M. Prentiss, Phys. Rev. Lett. 95 (2005) 243601.
%
\bibitem{agarwal}
G. S. Agarwal, and S. Dasgupta, Phys. Rev. A 65 (2002) 053811.
%
\bibitem{zhang}
H. R. Zhang, L. Zhou and C. P. Sun,
Phys. Rev. A  80 (2009) 013812.
%
\bibitem{naraschewski}
M. Naraschewski, D.M. Stamper-Kurn, Phys. Rev. A 58 (1998) 2423.
%
\bibitem{dutton} Z. Dutton, and L. V. Hau, Phys. Rev. A 70 (2004)
053831.
%
\bibitem{pethick}
C. J. Pethick and H. Smith, Bose-Einstein Condensation in Dilute
Gases, Cambridge, Cambridge, 2002.
%
\bibitem{eit3}
M.O. Scully, M.S. Zubairy, Quantum Optics, Cambridge, Cambridge, 1997.
%
\bibitem{Yariv}
A. Yariv, Optical Electronics, Saunders College,Holt,Rinehart and Wiston, 1991.
%
\bibitem{Alphan}
A. Sennaroglu, Photonics and Laser Engineering: Principles,
Devices and Applications, New Yowk, McGraw-Hill, 2010.
%
\bibitem{Yariv1}
L. Casperson, A. Yariv, Appl. Phys. Lett. 12 (1968) 355.

\end{thebibliography}
\end{document}